\documentclass{article}
\usepackage {graphicx}

\begin{document}

\bigskip

\begin{center}
{\large \bf Lithium Experiment in the Interplay of Solar and
Neutrino Physics.}

\vskip 0.3in

Anatoly Kopylov and Valery Petukhov\\ Institute of Nuclear
Research of Russian Academy of Sciences \\ 117312 Moscow, Prospect
of 60th Anniversary of October Revolution 7A
\end{center}

\begin{abstract}
In a future study of solar neutrinos a special emphasis should be
given to the measurement of the fluxes of neutrinos generated in
CNO cycle, because this is a direct way to measure with record
uncertainty -- less than 1\% the contribution to the solar
luminosity of a pp-chain of reactions. Combined with the
luminosity constraint method suggested by M. Spiro and D. Vignaud
this will put the foundation for the further substantial progress
in the study of neutrinos and thermonuclear reactions in the
interior of the Sun. So far the hypothesis of CNO cycle hasn't
found any experimental confirmation. A lithium-based radiochemical
detector has a potential to detect neutrinos from CNO cycle what
will be a direct proof of its existence. This will be a stringent
test of the theory of stellar evolution and will complete a
long-standing goal of recording the neutrino spectrum of the Sun.
The energy generated in the Sun and neutrino oscillation
parameters are inherently interconnected in this research. The
analysis shows that although a lithium detector is a radiochemical
one, principally it is possible to find separately the fluxes of
$^{13}$N- and $^{15}$O-neutrinos from the total neutrino capture
rate measured by this detector.
\end{abstract}

For nearly 40 years the solar neutrino problem, born after the
first results of the chlorine experiment \cite{1} has been
exciting the interest of a scientific community to the solar
neutrinos. The remarkable progress achieved in a number of
experiments with solar neutrinos \cite{2} with a culmination of
KamLAND \cite{3} has shown unambiguously that solar neutrinos do
oscillate and the parameters of neutrino oscillations responsible
for this belong to the MSW LMA region \cite{4}, which is now split
into two sub-regions so that at $3\sigma$ we have

\begin{center}
$5.1 \times 10^{-5} eV^2 < \Delta m^{2} < 9.7 \times 10^{-5}
eV^{2}$
\end{center}

\begin{center}
$1.2 \times 10^{-4} eV^{2} < \Delta m^{2} < 1.9 \times 10^{-4}
eV^{2}$
\end{center}

\noindent and for a mixing angle $\theta _{\odot}$

\begin{center}
$0.29 < \tan^{2} \theta _{\odot} < 0.86$
\end{center}

\noindent The peculiar thing is that probably the most decisive
result about oscillation of solar neutrinos was obtained in
experiment (KamLAND) dealing not with solar neutrinos but with
antineutrinos from reactors.

The further progress can be achieved by increasing the accuracy of
measurements of neutrino fluxes. Here some new aspect has arisen,
which is connected with the possibility to increase drastically
the accuracy in the evaluation of the contribution of pp chain to
the total luminosity of the Sun. The thing is that, as it is
known, there are basically two sources of solar energy: the
pp-chain of reactions and CNO-cycle, the latter is presented on
Fig.1 \cite{5}.

\begin{figure}[!ht]
\centering
\includegraphics[width=3in]{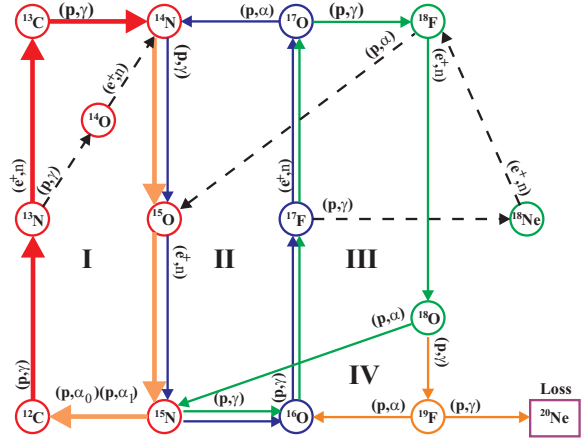}
\caption{The CNO cycle of thermonuclear reactions.}
\end{figure}
The CNO cycle contributes to the solar luminosity 1.5\% according
to a standard solar model (SSM) BP2000 \cite{6}. But so far
there's no experimental evidence of its existence. The present
limit for the contribution of a CNO cycle to the solar luminosity
is 7.3\% \cite{7}. It means that although the luminosity of the
Sun is measured with the accuracy better than 1\% one can not use
this high accuracy in the evaluation of the physical quantities
because the lack of the data on CNO neutrinos is smudging the
picture. To eliminate it one is obliged to remove the uncertainty
of CNO cycle and the only way to do it is to measure the fluxes of
neutrinos from $^{13}$N and $^{15}$O decays. To reach this aim it
would be adequate to measure the neutrinos from $^{13}$N and
$^{15}$O with the accuracy of about 20\% . This will enable to
determine the contribution of CNO cycle to the solar luminosity as
the value of about 1.5 $\pm$ 0.3 \% (if to take the number
predicted by SSM). Then the contribution of pp-chain to the solar
luminosity will be determined with the unprecedented accuracy as
the value 98.5 $\pm$ 0.3 \% (here we neglect the small corrections
due to the gravitational energy change during the current
evolution of the Sun which is calculated to be in module less than
0.1\% according to BP2000). From this one can envisage the new
perspectives for the physics of solar interior and for finding the
mixing angle $\theta _{\odot}$ by means of a luminosity constraint
\cite{8} applied to neutrino fluxes of pp-chain as the main source
of a solar energy. In the future the new experiments will measure
with much better accuracy, on the level of a few percent, the
fluxes of pp and $^7$Be neutrinos, which determine the major
energy production of the Sun. Thus for a further progress in the
study of solar neutrinos it is vital to measure the capture rate
by a lithium target. It was shown in \cite{9} that principally it
is even possible to find separately the fluxes of $^{13}$N and
$^{15}$O neutrinos although a lithium detector is a radiochemical
one, i.e. it measures the total capture rate from all neutrino
sources. In fact, the situation in this aspect is quite good for a
lithium target because due to a relatively high threshold of this
detector the rate from $^{13}$N neutrinos is much less than the
one from $^{15}$O neutrinos: the ratio is approximately 1/5, see
Table 1 taken from ref.6.
\begin{center}
\begin{table}[!ht]
Table 1. Standard Model Predictions (BP2000): solar neutrino
fluxes\\ and neutrino capture rates, with 1$\sigma$ uncertainties
from all sources\\ (combined quadratically).\\
\begin{tabular}
{|c|c|c|c|c|} \hline Source& Flux \par
(10$^{10}$cm$^{-2}$s$^{-1}$)& Cl
\par (SNU)& Ga \par (SNU)& Li \par (SNU) \\ \hline pp&
5.95(1.00$^{+0.01}_{-0.01}$)& 0.0& 69.7& 0.0
\\ \hline pep& 1.40$\times $10$^{-2}$(1.00$^{+0.015}_{-0.015}$)& 0.22& 2.8& 9.2
\\ \hline hep& 9.3$\times $10$^{-7}$& 0.04& 0.1& 0.1
\\ \hline $^{7}$Be& 4.77$\times $10$^{-1}$(1.00$^{+0.10}_{-0.10}$)& 1.15& 34.2& 9.1
\\ \hline $^{8}$B& 5.05$\times $10$^{-4}$(1.00$^{+0.20}_{-0.16}$)& 5.76& 12.1& 19.7
\\ \hline $^{13}$N& 5.48$\times $10$^{-2}$(1.00$^{+0.21}_{-0.17}$)& 0.09& 3.4& 2.3
\\ \hline $^{15}$O& 4.80$\times $10$^{-2}$(1.00$^{+0.25}_{-0.19}$)& 0.33& 5.5& 11.8
\\ \hline $^{17}$F& 5.63$\times $10$^{-4}$(1.00$^{+0.25}_{-0.25}$)& 0.0& 0.1& 0.1
\\ \hline Total& & 7.6$^{+1.3}_{-1.1}$& 128$^{+9}_{-7}$& 52.3$^{+6.5}_{-6.0}$
\\ \hline
\end{tabular}
\end{table}
\end{center}
This helps in the interpretation of the data because the
uncertainty of the rate from CNO neutrinos depends mainly upon the
uncertainty of the $^{15}$O neutrino flux. But still it is
possible to resolve these two sources. This will be useful for
getting the abundances of $^{12}$C and $^{14}$N in the central
zone of the Sun, which have rather peculiar distribution as the
calculation by the SSM reveals, see Fig.2 drawn from the
calculated data presented in ref.10. The information about how
well the SSM describes the role of CNO cycle in the Sun will be
very useful for understanding other stellar objects, primarily
those where CNO-cycle plays a major role.
\begin{figure}[!ht]
\centering
\includegraphics[width=3in]{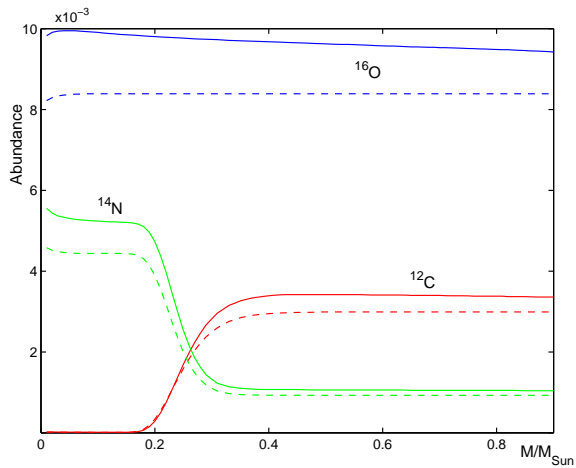}
\caption{The abundances of $^{12}$C and $^{14}$N in the interior
of the Sun.}
\end{figure}

Now, let's suppose the capture rate by a lithium detector has been
measured. What can be a procedure for the evaluation of the fluxes
of neutrinos from CNO cycle and of the mixing angle $\theta
_{\odot}$? By that time $\Delta m^2$ will be measured by KamLAND
with a very good accuracy, so let's take it as a known value. Then
as a first step we should subtract from the total measured capture
rate the effect from neutrinos $^7$Be, pep and $^8$B. By the time
we get the data in a lithium experiment these fluxes will be known
with good accuracy (with the possible exception of the
pep-neutrinos, but this is of little consequence because the ratio
of the pep flux to the pp flux is well known). In fact, to
subtract the effect from all these sources we don't need
extraordinary accuracy, because our aim is to get the effect from
CNO neutrinos with the accuracy of about 20\% what is enough as we
have seen earlier. To do this we need the accuracy in the
evaluation of the pep, $^7$Be and $^8$B fluxes of about 6-8\% . To
find the contribution of beryllium neutrinos one should know not
only the flux of beryllium neutrinos but also the shape of the
energy spectrum of these neutrinos due to thermal broadening. The
details of this were discussed in \cite{11}. The point is that in
the laboratory conditions the $^7$Be line will not produce $^7$Be
on lithium since the reaction of $^7$Be production is reverse to
electron capture by $^7$Be. If to consider electron screening in
terrestrial atoms, the energy of beryllium line is even lower than
a threshold for $^7$Be production. But in the Sun high temperature
produces the thermal broadening of the $^7$Be line, as it was
discussed in \cite{12} and later was computed with high accuracy
by Bahcall \cite{13}. Because of this some fraction of the line
with the energy higher than the threshold will produce $^7$Be. The
effect is model dependent. The fact that the measured flux of
boron neutrinos is in a good agreement with the one predicted by
BP2000 shows that the model gives the correct temperature map of
the interior of the Sun, hence there is good reasoning to believe
that the thermal broadening of the beryllium line is described by
the model correctly. The substantial issue is that while the
contribution of CNO cycle to the solar energy is only 1.5\% , the
weight of neutrinos from CNO cycle in the production rate of
$^7$Be on $^7$Li is about 30\% , so that for the total capture
rate expected for a lithium target 23 SNU neutrinos from CNO cycle
contribute 7 SNU. If we take the parameters of neutrino
oscillations of the best-fit point, make the estimates for a
detector with 10 tons of lithium and take pure statistical
uncertainties as it was done in \cite{9}, then the capture rate on
a lithium target can be measured with the accuracy of
approximately 1 SNU for 16 Runs total performed during 4 years of
measurements. After subtracting the rate from these three sources
of a pp-chain one gets the rate from neutrinos of CNO cycle. Now
we have two possibilities: first, we can take the ratio of
$^{13}$N to $^{15}$O neutrinos as a given one by a SSM, or we can
find separately the contribution of these two neutrino sources to
the total capture rate solving the system of two equations:

$$ \left\{
\begin{array}{rcl}
L_H+L_{CN}+L_{NO}&=&L_{\odot}\\ R_H+R_{CN}+R_{NO}&=&R_{Li}\\
\end{array}
\right.
 $$

\noindent Here $L_{H}$, $L_{CN}$ and $L_{NO}$ are the
contributions to the solar luminosity of pp-chain and two
half-cycles of CNO cycle, $R_{Li}$ , $R_{H}$ - the measured and
estimated for the hydrogen sequence rates in a lithium detector,
$R_{CN}$ and $R_{NO}$ -- the rates from neutrinos born in
$^{13}$N- and $^{15}$O-decays, R = yL/4$\pi
$r$_{SUN}^{2}\varepsilon $, where r$_{SUN}$ -- the distance from
Sun to Earth, $\varepsilon $ is the energy contributed to the Sun
per one neutrino emitted in each half-cycle of CNO-cycle and y --
the capture rate per one neutrino of $^{13}$N- or
$^{15}$O-spectra. One can see from these equations that
principally it is possible to find separately the fluxes of
$^{13}$N and $^{15}$O neutrinos. Let's look more in the details.
If the fluxes of neutrinos from a hydrogen sequence are measured
with very good accuracy then the only unknown thing is the energy
of CNO cycle. But the energy generation in this cycle proceeds by
two half-cycles: from $^{12}$C to $^{14}$N (first one) and from
$^{14}$N to $^{12}$C (second one). The rates depend upon the
abundance of $^{12}$C and $^{14}$N in the interior of the Sun.
This is why the fluxes of $^{13}$N and $^{15}$O neutrinos are
different, see Table1. Figure 2 shows the distribution of the
abundance of $^{12}$C and $^{14}$N along the profile of the Sun
(in the mass ratio units) \cite{10}. One can see that the center
of the Sun is depleted by $^{12}$C (it is burned out) while is
enriched by $^{14}$N (it is accumulated). The question is: will
this abundance distribution be confirmed by experiment? For the
first half-cycle the energy release $E_1 = ^{12}$C$ + 2^1$H$ -
^{14}$N. For the second one $E_2 = ^{14}$N$ + 2^1$H$ - ^{12}$C$ -
\alpha $. The total energy release will be as it is well know $E_1
+ E_2 = 4^1$H$ - \alpha $. The energy released in the first
half-cycle is a bit smaller than the one in the second half-cycle
$E_2 - E_1 = 2 (^{14}$N$ - ^{12}$C$) - \alpha $. It is about 3.3
MeV. And if to take into account that the energy of neutrino
emitted in the first half-cycle is less that the energy of the
neutrino in the second one, we obtain that the in the first
half-cycle the Sun gets less energy only by about 3.1 MeV than in
the second half-cycle, this means that these energies are very
close. What concerns the rate of a lithium detector, the situation
here is very different. The contribution of the $^{15}$O-neutrino
is 5 times bigger than the one of $^{13}$N-neutrinos. In other
words, the straight lines corresponding to equations (1) and (2)
of the system are not parallel and the system of equations has a
solution. Now, having found the CNO neutrinos, we have found the
energy produced in CNO cycle and subtracting this energy from the
solar luminosity we find the contribution to the solar luminosity
of a pp-chain. Then we can use the luminosity constraint method.
At this phase it is important to have the accuracy of the
measurements of neutrino fluxes of a pp-chain, primarily, of the
flux of pp-neutrinos, as high as possible. Taking as the
experimental points the ones simulated by Monte-Carlo using the
error bars given by experiments we obtain a new
$\theta^{\prime}_{\odot}$ corresponding to each Monte-Carlo
simulation. As a result we have found the field of points with the
corresponding weights in this Monte-Carlo simulation. Obviously
the input $\theta _{\odot}$ and output $\theta^{\prime}_{\odot}$
should coincide within the experimental uncertainties as it is
illustrated on Fig.3.

\begin{figure}[!ht]
\centering
\includegraphics[width=2in]{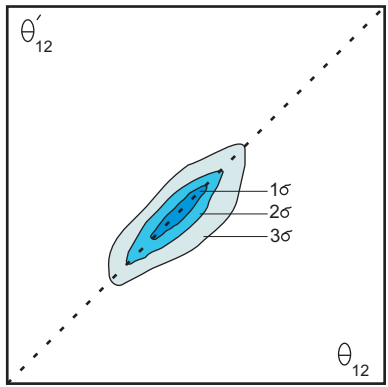}
\caption{The $\theta $-diagram.}
\end{figure}

For each confidence level one can find the corresponding allowed
region for $\theta _{\odot}$ which is located in the vicinity of a
diagonal of Fig.3. This is a general procedure, which can be used
if the rate has been measured by a lithium detector. The useful
thing is that as a result of this procedure we can get both the
energy produced in CNO cycle and the mixing angle $\theta
_{\odot}$ and there's a close interconnection between these two
values. This is a good example of the interplay of a solar and
neutrino physics apart from the luminosity constraint method by
itself.

The last issue we would like to address in this paper is the
question: is it really very interesting to measure precisely the
mixing angle $\theta _{\odot}$? It was proposed in \cite{14} to
measure tan$^2 \theta _{\odot}$ with the accuracy of about 14\% in
a dedicated reactor antineutrino experiment with a base of about
70 km oriented to the minimum of the oscillation curve. This is,
of course, a very good idea. But the Sun can give us more. The
accuracy of determining $\theta _{\odot}$ in Sun's laboratory is
limited mainly by the accuracy of measurement of neutrinos of
pp-chain. So here we can go further, till the level of a few
percent. The reward would be not only a better accuracy in finding
a mixing angle, this is just the way of searching for something
new by means of precise measurements as it has been done by
Raymond Davis when he decided to measure the flux of boron
neutrinos from the Sun. The potentiality of this research is still
high, starting from more precise understanding of the interior of
the Sun, continuing to the study of neutrino physics and, as it
well may be, till the possible discovery of the unknown source of
solar energy.

To summarize we should note that a lithium experiment is the only
way to find with unprecedented accuracy the energy produced by a
pp-chain of reactions in the interior of the Sun by means of
measuring the fluxes of neutrinos generated in a CNO cycle. This
information is vital for further study of the thermonuclear
processes in the interior of the Sun and can be effectively used
for finding the mixing angle $\theta _{\odot}$ with the accuracy
of $tan^2 \theta _{\odot}$ on the level of a few percent. The
study of CNO is very important also as a precise test of the
theory of stellar evolution.

This work was supported in part by the Russian Fund of Basic
Research, contract N 01-03-16167-A and by the grant of Russia
``Leading Scientific Schools'' LSS-1782.2003.2. The authors deeply
appreciate the very stimulating discussions with G.Zatsepin,
L.Bezrukov, V.Kuzmin, Yu.Efremenko, G.Seidel, R.Lanou, M.Nakahata,
S.Turch-Chieze.


\begin{thebibliography}{99}

\bibitem{1} R.Davis Jr., J.C.Evans, V.Radeka, and L.Rogers, 1972, {\it Neutrino 72 Conference,
Balatonfured, Hungary \bf 1} 5-27

\bibitem{2} B.T.Cleveland et al., 1998, {\it Astrophys. J. \bf 496},
505\\ J.N.Abdurashitov et al., 2002, {\it J. Exp. Theor. Phys. \bf
95}, 181\\ W.Hampel et al. (GALLEX collaboration), 1999, {\it
Phys. Lett. \bf B447}, 127\\ T.Kirsten, 2002, {\it talk at the
XXth Int. Conf. On Neutrino Physics and Astrophysics (NU2002)},
Munich, May 25-30\\ M.Altmann et al., 2000, {\it Phys. Lett. \bf
B490}, 16\\ E.Bellotti et al. (GNO collaboration), 2000, in {\it
Neutrino 2000, Proc. of the XIXth Int. Conf. On Neutrino Physics
and Astrophysics},\\ 16-21 June, {\it eds. J.Law}\\ R.W.Ollerhead,
J.J.Simpson, 2001, {\it Nucl. Phys. B (Proc. Suppl.) \bf 91}, 44\\
Y.Fukuda et. al., 1996, {\it Phys. Rev. Lett. \bf 77}, 1638\\ S.
Fukuda et al., 2001, {\it Phys. Rev. Lett. \bf 86}, 5651\\
Q.R.Ahmad et al., 2001, {\it Phys. Rev. Lett. \bf 87}, 71301\\
Q.R.Ahmad et al., 2002, {\it Phys. Rev. Lett. \bf 89}, 11301

\bibitem{3} K.Eguchi et al. (KamLAND collaboration), {\it
hep-ex/0212021}

\bibitem{4} V.Barger, D.Marfatia, {\it hep-ph/0212126}\\
G.L.Fogli, E.Lisi, A.Marrone, D.Montanino, {\it hep-ph/0212127}\\
M.Maltoni, T.Schwetz, J.W.F.Valle, {\it hep-ph/0212129}\\
A.Bandyopadhyay, S.Choubey, R.Gandi, S.Goswami, {\it
hep-ph/0212146}\\ J.N.Bahcall, M.C.Gonzalez-Garchia,
C.Pe\~{n}a-Garay, {\it hep-ph/0212147}\\ P.C.de Hollanda and
A.Yu.Smirnov, {\it hep-ph/0212270}

\bibitem{5} E.C.Adelberger at al., {\it astro-ph/9805121}

\bibitem{6} J.N.Bahcall, M.N.Pinsonneault, S.Basu, {\it
astro-ph/0010346}

\bibitem{7} J.N.Bahcall, M.C.Gonzalez-Garcia, and C.Pe\~{n}a-Garay, {\it
hep-ph/0212331}

\bibitem{8} M.Spiro and D.Vignaud, 1990, {\it Physics Letters B, \bf 242} 279-284

\bibitem{9} A.Kopylov, V.Petukhov, {\it hep-ph/0301016}

\bibitem{10} J.N.Bahcall, 1989, {\it Neutrino Astrophysics} Cambridge University Press,
{\it Cambridge}

\bibitem{11} J.N.Bahcall, 1994, The $^7$Be Solar Neutrino Line: A Reflection of the Central
Temperature Distribution of the Sun, {\it Preprint IASSNS-AST
93/40}

\bibitem{12} G.V.Domogatsky, 1969, {\it Preprint of Lebedev Phys. Inst., Moscow,
no.153}

\bibitem{13} J.N.Bahcall, 1978, {\it Rev.Mod.Phys. \bf 50}, 881

\bibitem{14} A.Bandyopadhyay, S.Choubey, S.Goswami, {\it
hep-ph/0302243}

\end{thebibliography}
\end{document}